\documentclass[aps,prb,twocolumn,showpacs]{revtex4}
\usepackage{graphicx}

\begin{document}

\title{Vortex images on Ba$_{1-x}$K$_x$Fe$_2$As$_2$ observed directly by the magnetic force microscopy}

\author{Huan Yang,$^{1,*}$ Bing Shen,$^2$ Zhenyu Wang,$^2$ Lei Shan,$^2$ Cong Ren,$^2$ and Hai-Hu Wen$^{1,\dag}$}

\affiliation{$^1$Center for Superconducting Physics and Materials, National Laboratory of Solid State Microstructures and Department of Physics, Nanjing University, Nanjing 210093, China}
\affiliation{$^2$Institute of Physics and Beijing National Laboratory for Condensed Matter Physics, Chinese Academy of Sciences, P.O. Box 603, Beijing 100190, China}

\begin{abstract}
The vortex states on optimally doped Ba$_{0.6}$K$_{0.4}$Fe$_2$As$_2$ and underdoped
Ba$_{0.77}$K$_{0.23}$Fe$_2$As$_2$ single crystals are imaged by
magnetic force microscopy at various magnetic fields below 100$\;$Oe.
Local triangular vortex clusters are observed in optimally doped
samples. The vortices are more ordered than those in
Ba(Fe$_{1-x}$Co$_{x}$)$_{2}$As$_{2}$, and the calculated pinning
force per unit length is about 1 order of magnitude weaker than
that in optimally Co-doped 122 at the same magnetic field,
indicating that the Co doping at the Fe sites induces stronger
pinning. The proportion of six-neighbored vortices to the total
amount increases quickly with increasing magnetic field, and the
estimated value reaches 100\% at several tesla. Vortex chains are
also found in some local regions, which enhance the pinning force
as well as the critical current density. Lines of vortex chains
are observed in underdoped samples, and they may have originated
from the strong pinning near the twin boundaries arising from the
structural transition.
\end{abstract}

\maketitle

\section{Introduction}
Since the discovery of the iron-based superconductors,\cite{FeAs}
the mechanism of their superconductivity and vortex dynamics has
attracted considerable interest. Multiple electron and hole Fermi
pockets, as well as multiple superconducting gaps, have greatly enriched
the physics of superconductivity in this new
system.\cite{ARPES1,ARPES2} Theoretically it was suggested that
the unique sign-reversal $s$-wave pairing, namely, $s_{\pm}$, could
be the main pairing symmetry of the iron pnictide superconductors,
and the nesting between hole and electron pockets is important for
achieving superconductivity.\cite{s+-} This extended $s$-wave
model results in nodeless superconducting gaps and a sign change
of the order parameter between the nested pockets, which seems to
be supported by scanning tunneling microscopy measurement in
Fe(Se,Te) samples.\cite{HanaguriSTM} Recent angle-resolved
specific heat measurements show a fourfold oscillation of the
specific heat as a function of the in-plane magnetic field
direction, which suggests that the gap is
anisotropic.\cite{WenNComm} As a type-II superconductor with such a
multiband and fascinating pairing symmetry, the vortex dynamics of
pnictides is also attractive. The 122 family of iron pnictides
provides a good opportunity to explore the vortex dynamics because
of the availability of its high-quality single crystals. The
parent compound BaFe$_2$As$_2$ has both hole and electron pockets
with almost balanced charge carriers. Superconductivity can be
achieved via chemical doping, for example via K substitution at Ba
sites\cite{BaKFeAs} and Co substitution at Fe
sites.\cite{BaFeCoAs} The multiband property plays an important
role in electric transport for both hole- and electron-doped
samples.\cite{BaKRT,CoRT} The magnetization
measurements\cite{BaKMH,CoMH} show that the hole or electron
optimally doped samples both have the second magnetization peak
effect and a very similar vortex phase diagram. For the vortex
imaging measurement, almost all the measurements were taken on
Ba(Fe$_{1-x}$Co$_{x}$)$_{2}$As$_{2}$ or
Ba(Fe$_{1-x}$Ni$_{x}$)$_{2}$As$_{2}$ samples. Different detecting
methods give similar results, i.e., the vortex structure seems
to be very disordered because of the strong
pinning.\cite{VortexImage1,VortexImage2,VortexImage3,VortexImage4,VortexImage5}
Recently, the scanning tunneling microscopy measurement on
Ba$_{0.6}$K$_{0.4}$Fe$_2$As$_2$ shows the ordered vortices as well
as the Andreev bound states,\cite{ShanNPhy} which seems to be very
different from Co-doped 122 in which neither the ordered vortex
lattice or the in-core Andreev bound states were
observed.\cite{VortexImage4} This is quite natural since the
Co doping takes place right at the Fe-As planes, while the
K doping at the Ba sites induces most probably the off-plane
disorders. Therefore it is very interesting to investigate the vortices
at low fields on K-doped 122 samples and compare them with those in the
Co-doped samples. In this paper we present the direct imaging of
vortices on high-quality Ba$_{1-x}$K$_x$Fe$_2$As$_2$ single
crystals detected by  magnetic force microscopy below
100$\;$Oe. The difference in vortex structure in both K-doped and
Co-doped samples is analyzed and discussed in detail.

\section{Experiments}

The Ba$_{1-x}$K$_x$Fe$_2$As$_2$ single crystals were grown by the
self-flux method using FeAs as flux, and the detailed
procedures of synthesizing are similar to previous
reports.\cite{sample1,sample2,ShanNPhy} The measurements of x-ray
diffraction indicate a highly $c$-axis orientation and crystalline
quality of our samples. The bulk diamagnetic characterizations of
single crystals were measured by a magnetic property measurement
system (MPMS, Quantum Design). Magnetic force microscopy (MFM)
used in this work is the atto-MFM system (attocube) based on the
physical properties measurement system (PPMS-9, Quantum Design).
Hard magnetic coating point probes from NanoWorld were used for
all the measurements. The vortex figures are made by  WSxM
software.\cite{WSxM} For every MFM measurement, the
Ba$_{1-x}$K$_x$Fe$_2$As$_2$ single crystal was mounted on the
sample holder of MFM immediately after it was cleaved along the
$ab$ plane in air at room temperature. The fresh top surface
is always flat and mirror-like for the MFM measurements, and usually
the measured roughness on the cleaved surface is less than 1$\;$nm,
which approaches the measuring precision of the system. Then the
sample was cooled in a low-pressure helium gas environment.
The magnetic property and the MFM measurements were carried out
with the magnetic field perpendicular to the top surface ($ab$
plane). The magnetic field was applied at the temperature above
critical temperature to obtain the field-cooling process for
MFM measurements. The attocube scanners in the atto-MFM system were
calibrated by a standard sample to obtain the exact scanning
parameters at different temperatures. The magnets of the MPMS and
PPMS systems were degaussed before the measurements to minimize
the residual magnetic field. The first step in the measurement is
to find a rather flat place by a tip tapping mode. Then we
keep a constant distance between the tip and the sample surface (e.g., 10$\;$nm)
and detect the resonance frequency change in the presence of the field
distribution around the vortices. Since the density of vortices
changes with the magnetic field, we use more scanning pixels to
get a clearer image at higher fields.

\section{Results}

\subsection{Sample characterization}

\begin{figure}
\includegraphics[width=8cm]{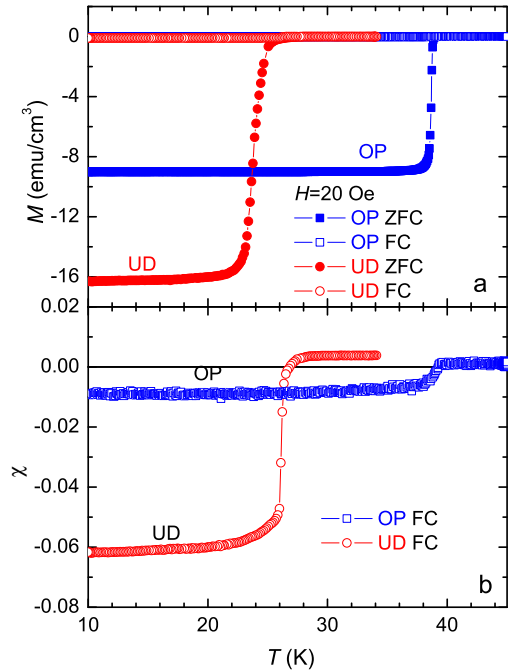}
\caption{(Color online) (a) Temperature dependence of volume
magnetization of the optimally doped
Ba$_{0.6}$K$_{0.4}$Fe$_2$As$_2$ and underdoped
Ba$_{0.77}$K$_{0.23}$Fe$_2$As$_2$ samples after zero-field cooling
(ZFC) and field cooling (FC) at 20$\;$Oe. The difference in the
ZFC magnetization of the two samples comes from the different
demagnetizing effect. (b) Field cooled susceptibilities versus
temperature. Since the susceptibilities are rather small, it seems
that very few vortices are excluded from the sample in the FC
process.} \label{fig1}
\end{figure}

Figure \ref{fig1}(a) shows the temperature dependence of the
volume magnetization (M) after zero-field-cooling (ZFC) and
field cooling (FC) processes. Both the optimally doped (OP)
sample and underdoped (UD) sample used in our MFM measurements
shows very good superconducting transitions. The critical
transition temperature of the OP sample is 38.8$\;$K
($10\%M_{T=10\;\mathrm{K}}$) with a transition width of 0.6$\;$K,
while the value for the UD sample it is 24.7$\;$K with a transition
width of 1.5$\;$K. Ba$_{1-x}$K$_x$Fe$_2$As$_2$ single crystals are
usually very thin, so the demagnetizing factor approaches 1.0.
That is the reason the ZFC magnetization values of these two
samples are different. The error in the measurement of
dimensions, especially the thickness, could give error in the
calculation of the ZFC susceptibility. Figure~\ref{fig1}(b) shows
the temperature dependence of the FC volume susceptibility
($\chi_\mathrm{FC}$), from which we can estimate the ratio of
vortices excluded from the sample. The $\chi_\mathrm{FC}$ values of the OP
and UD samples are only 1\% and 6\%, which means that a large
number of vortices are pinned in the samples after field
cooling. In our MFM measurements we also find that the density of
vortices is close to that calculated from the magnetic field.

\subsection{Vortex image in OP samples}\label{OP}

\begin{figure}
\includegraphics[width=9cm]{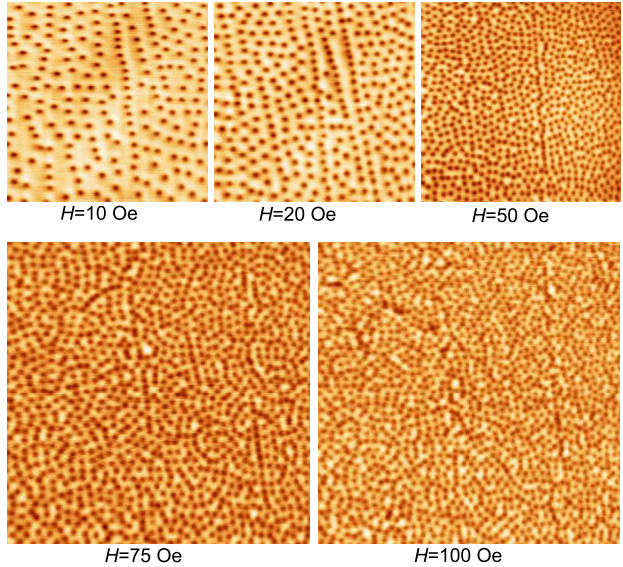}
\caption{(Color online) Vortex image of an OP
Ba$_{0.6}$K$_{0.4}$Fe$_2$As$_2$ single crystal measured by MFM at
2$\;$K and different magnetic fields from 10 to 100$\;$Oe.
The scanning range for each image is 19$\;\mu$m$\times19\;\mu$m.}
\label{fig2}
\end{figure}

In Fig.~\ref{fig2} we show the vortex image on the OP sample at
different magnetic fields at 2$\;$K by an FC process. The maximum
field reaches 100$\;$Oe which is almost the limit to distinguish
the nearest vortices in MFM measurement. The calculated number of
vortices in this certain range was almost the same as the
calculated magnetic flux at each field, which is consistent with
the FC susceptibility mentioned above. The distance between the
neighboring vortices seems to be very uniform, which is similar to
the Bitter decoration result on some
Ba(Fe$_{1-x}$Co$_{x}$)$_{2}$As$_{2}$ samples,\cite{VortexImage2}
but more ordered than other reports.\cite{VortexImage1,VortexImage3} Local vortex chains
observed in the vortex image are discussed in Sec.~\ref{UD}.

\begin{figure}
\includegraphics[width=6cm]{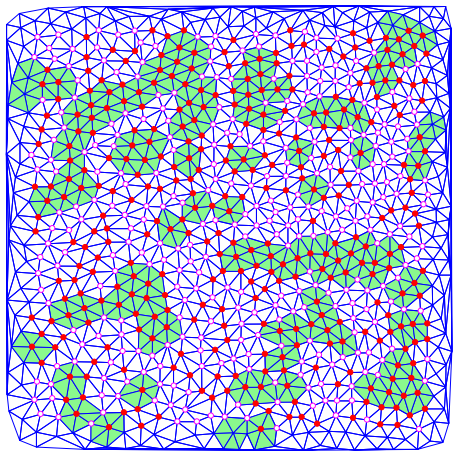}
\caption{(Color online) Delaunay triangulation of the vortices on the
OP sample at 100$\;$Oe and 2$\;$K. The red solid circles denote
6-neighbored vortices, while the pink empty ones denote the other
number of neighbored vortices. To avoid the error from the
vortices near the scanning edge, only the ones with the distance
more than 1$\;\mu$m from the edges are taken into account. The
light green blocks show area of distorted triangle lattice.}
\label{fig3}
\end{figure}

\begin{figure}
\includegraphics[width=8cm]{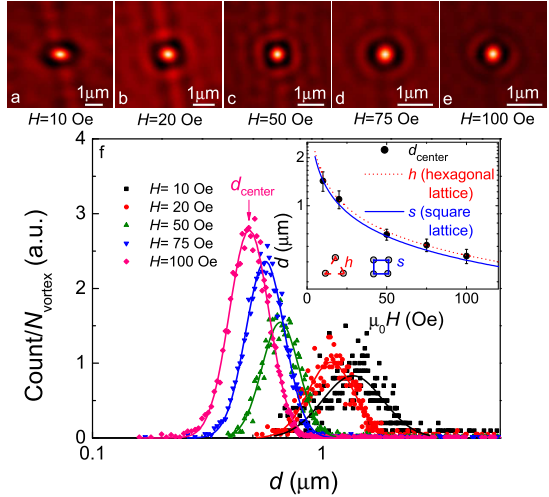}
\caption{(Color online) (a)-(e) Self-correlations of vortex images
at different fields. (f) Statistics of distance between the nearest
vortices ($d$) in Delaunay triangulation at various fields. The
solid line in (f) is the Gaussian fit in semilogarithmic scale.
The inset of (f) shows the comparison between the peak-value
($d_\mathrm{center}$) and the distances expected for a
normal square ($s$) and a hexagonal ($h$) vortex lattices. The
errors of $d_\mathrm{center}$ are full width at 90\% maximum. }
\label{fig4}
\end{figure}

\begin{figure}
\includegraphics[width=7cm]{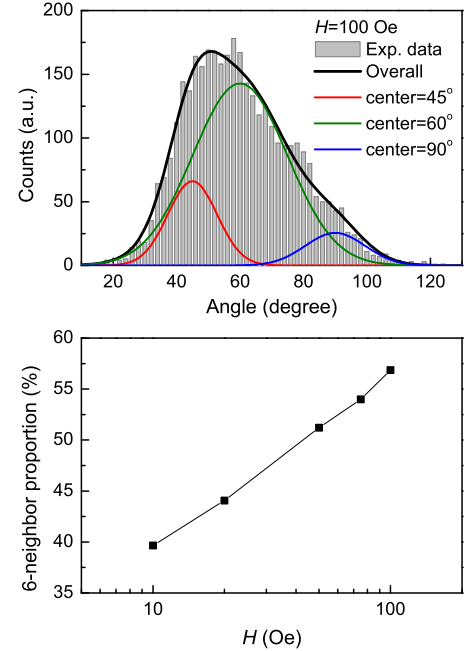}
\caption{(Color online) (a) Statistics on the angle values of
Delaunay triangles. The solid line shows the Gaussian fittings at
45, 90, and 60 degrees which are the typical angle of a square, or
a hexagonal vortex lattice. (b) The proportion of the 6-nearest
neighbored vortices at different fields in semi-log plot. It
is obvious that the vortices favor 6-neighbored more at a higher
fields. The proportion may reach 100\% at several tesla if we do linear
extrapolation in semi-log plot.} \label{fig5}
\end{figure}

To make the figure more clear, we took the coordinates of all the
vortex centers and used the Delaunay triangulation to figure out the
vortex distribution. The result at 100$\;$Oe is shown as an
example in Fig.~\ref{fig3}. One can find that almost half of the
vortices are six-nearest-neighbored, and more importantly there are
distorted triangle lattice fragments in some local areas. The
self-correlation figures for each vortex image at various fields
are shown in Figs.~\ref{fig4}(a)--(e). At 10$\;$Oe the irregular
loop around the center indicates that the nearest distance between vortices
has a broad distribution, and it means that the force among the
vortices is very small. The approximative four fold loop at 20$\;$Oe shows
that the vortices form a squarelike structure in the local area.
With increasing the magnetic field (or the density of vortices), the nearest pattern
loop in the self-correlation figure will change into a circular shape,
which suggests that pairs of neighbored vortices have contiguous distances
but random orientations. They are not ordered enough to form the vortex lattice.
Statistics of the distances between two nearest vortices by the Delaunay triangulation
method are shown in Fig.~\ref{fig4}(f). The Gaussian function fits the
statistic data very well in semilogarithmic scale, and the
maximum points from the fits are between the values calculated
from a normal square and a hexagonal vortex lattice at the same
fields. The half-width decreases quickly with increasing magnetic field, which means that the
stronger force at higher fields makes the distance between
nearest vortices more uniform. It is very difficult to get the square pattern in Delaunay
triangulation plots, so we do the statistic of angles of the
Delaunay triangles as shown in Fig.~\ref{fig5}(a). According to
the fitting to a Gaussian distribution, the square (with characteristic
angles of 45$^\circ$ and 90$^\circ$) and hexagonal structure (with a characteristic
angle of 60$^\circ$) vortices coexist at 100$\;$Oe. Figure~\ref{fig5}(b) shows
the field dependence of the number ratio of six neighbored vortices, which behaves
linearly in a semilog plot. The ratio increases with magnetic field, as the vortex system
favors a six-neighbored situation at high magnetic field. If we do the
linear extrapolation on the curve to higher fields, the ratio may
reach 100$\;\%$ at a field magnitude of several tesla, which means that almost all of the
vortices have six nearest neighbors at such high fields. So it is
not strange that the vortices become the ordered hexagonal phase
in several vortex spacings at a magnetic field of 9$\;$T.\cite{ShanNPhy}

\begin{figure}
\includegraphics[width=9cm]{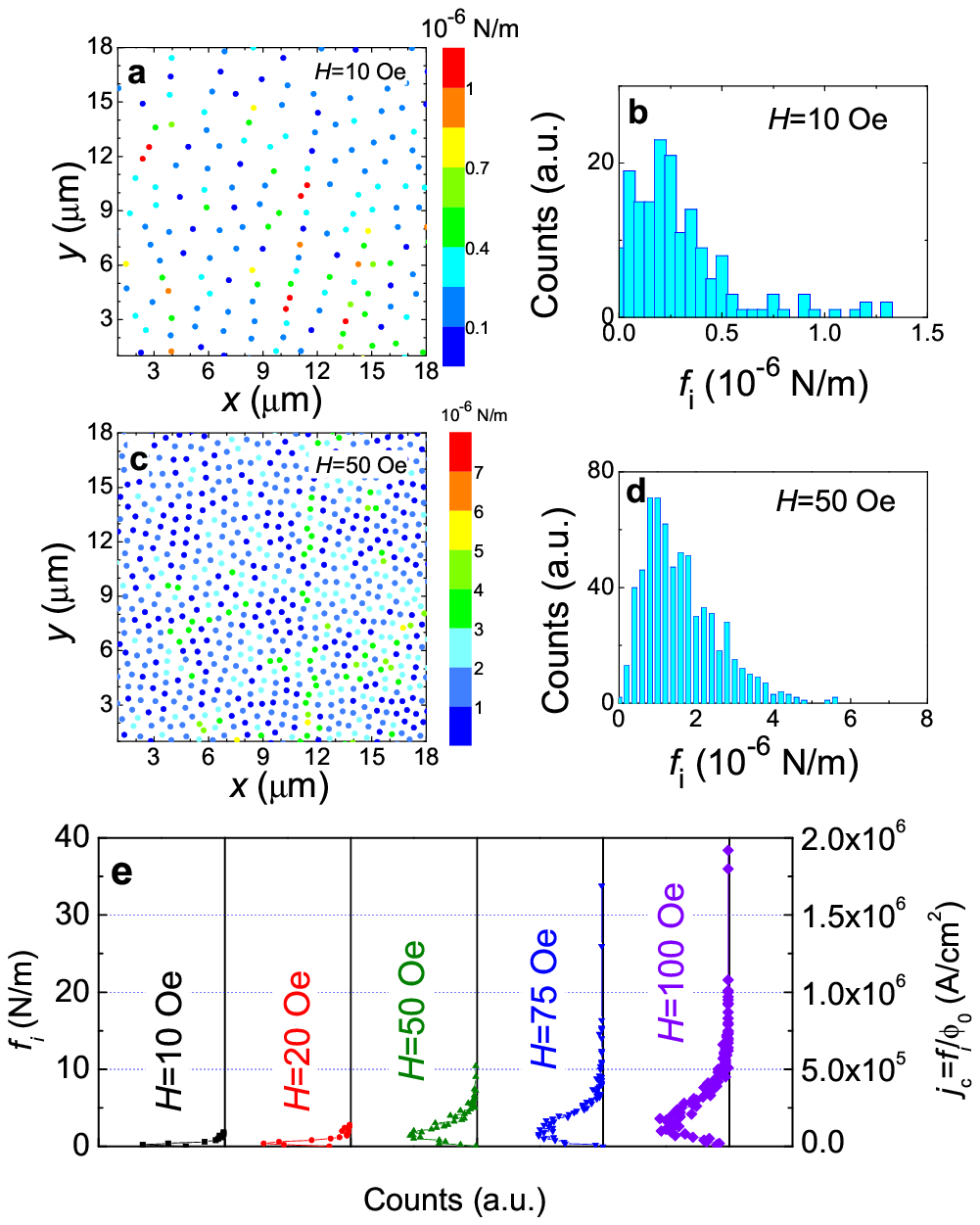}
\caption{(Color online) Two-dimensional color mapped pinning force per unit
length of every vortex at 10$\;$Oe (a) and 50$\;$Oe (c), while the
distributions of the pinning force per unit length are shown in
the histograms of (b) and (d), respectively. The highest frequency value
of the pinning force per unit length at 10$\;$Oe is about
$2\times10^{-7}\;$N/m, which is 1 order of magnitude smaller
than the value in the Co-doped 122 system at the same
field.\cite{VortexImage1} (e) Pinning force per unit length and
the related critical current density distributions at each
magnetic field.} \label{fig6}
\end{figure}

The vortices in the K-doped sample seem to distribute more orderly
than in the Co-doped sample, which requires some data to prove. The
pinning force per unit length and the pinning energy are the
parameters characterizing the vortex pinning. The vortex structure
will be more disordered if the pinning energy or the pinning force
is bigger. As a sum of the scalar quantities, the
pinning energy has a more close relationship with the magnitude of
the magnetic field than the pinning force, so we calculate the pinning
force per unit length of the vortices to do further
analysis. The pinning force per unit length on the $i$th
vortex, which is related to the shielding current at its location from the other
vortices, can be expressed as
\begin{equation}
\mathbf{f}_i=\sum_{j\neq i}\frac{\phi_0^2}{2\pi\mu_0\lambda^3}\frac{\mathbf{r}_{ij}}{\left|\mathbf{r}_{ij}\right|}K_1\left(\frac{\left|\mathbf{r}_{ij}\right|}{\lambda}\right).
\label{eq1}
\end{equation}
Here $\phi_0$ is the magnetic flux quantum, $\mu_0$ is the
permeability of the vacuum, $\mathbf{r}_{ij}$ is the space vector
between the $j$th and $i$th vortex, and
$K_1(r_ij/\lambda)$ is the first-order modified Bessel function.
Figure~\ref{fig6} presents the absolute value of pinning force per
unit length ($f_i$) distributions for every vortex at 10 and
50$\;$Oe. The statistic result shows that most of the values of $f_i$
are around $2\times10^{-7}\;$N/m at 10$\;$Oe, which is 1 order
of magnitude weaker than that in the Co-doped 122 system at the same
magnetic field.\cite{VortexImage1} The penetration depth used here
is 0.25$\;\mu$m at 2$\;$K from values of the tunnel diode
resonator technology\cite{pendepthProzorov} and
$\mu$SR.\cite{pendepthuSR} It should be noted that if we use the
larger penetration depth, i.e., 1.2$\;\mu$m, as used for the Co-doped
sample in Ref.~\cite{VortexImage1}, the calculated value of $f_i$
would become even 1 order of magnitude smaller. So the big
difference in pinning force in the electron- or hole-doped 122 system
is not from the different chosen values of penetration depth.
Small-size normal cores are the pinning centers in
Ba$_{0.6}$K$_{0.4}$Fe$_2$As$_2$, which may originate from the
local doping-induced disorders or some local magnetic
moments.\cite{BaKMH} Although the phase
diagrams\cite{CoRT,BaKPhaseD} of superconductivity are similar for
both doping sides, the electron doping induces impurity by
substituting the Fe sites with Co, which may be the source of the
extra pinning centers.\cite{BaKRT}

The estimated critical current density $j_\mathrm{c}$ is proportional to the
pining force per unit length, i.e., $j_\mathrm{c}=f/\phi_0$.
Weak pinning in Ba$_{0.6}$K$_{0.4}$Fe$_2$As$_2$ means a
very small critical current density, but it is not consistent with other experiment results.
One reliable explanation is that the K-doped sample has higher $T_\mathrm{c}$, and the pair-breaking
scattering in the Co-doped sample suppresses the superfluid density.
In this way the intrinsic critical current densities are very different in K-doped or Co-doped samples.
It should be noted that the compared pinning force mentioned above is the average one.
For example, the calculated average $j_\mathrm{c}=10^4\;$A/cm$^2$ is very small at
10$\;$Oe. As the penetration depth here (0.25$\;\mu$m) is much
smaller than the average distance (1.4$\;\mu$m) between the
neighboring vortices, the rare vortices have very small
interactions, which may be the reason for the small
$j_\mathrm{c}$. The peak value of $j_\mathrm{c}$ increases
slightly with the magnetic field, as shown in Fig.~\ref{fig6}(e).
At 100$\;$Oe $j_\mathrm{c}$ at the most frequent position is about
$1.4\times10^5\;$A/cm$^2$ and the maximum value reaches
$1.9\times10^6\;$A/cm$^2$, which is of the same amplitude as the
value taken on the magnetization curve.\cite{BaKMH}

\begin{figure}
\includegraphics[width=8cm]{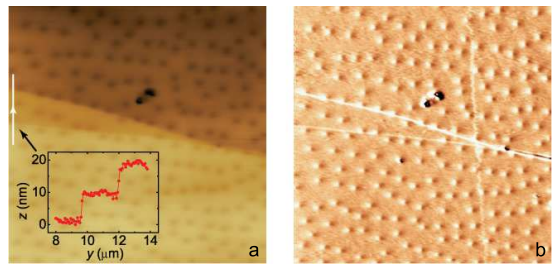}
\caption{(Color online) Vortex image (a) and its differential image (b) in the presence of two steps on the surface at 10$\;$Oe and 4.2$\;$K. The insert in (a) shows the height landscape along the white line. It is clear that the heights of the two steps are about 10$\;$nm, and the image dimension is 19$\;\mu$m$\times19\;\mu$m. The vortices near the steps were pinned along the edge of the steps. } \label{fig7}
\end{figure}

The newly cleaved fresh surface is very flat, except that some surface
steps are formed by the cleaving. In Fig.~\ref{fig7} we show the
case of two steps with about 10$\;$nm in height, and the vortices
were pinned by these steps. At first glance the pinning of
vortices by these steps is supposed to be induced by the
Bean-Livingston pinning.\cite{Bean-Livingston} When a vortex is
close to the parallel mirror surface, an attractive interaction is
formed between the vortex and its image (with opposite
sign).\cite{VortexImage1} In this case the vortices should stay at
the higher side of the stage. This kind of pinning can only happen
when the step is high enough leading to a large mirror area
parallel to the vortices. This can readily explain why the
vortices along the upper step locate on the high stage and keep
some distance away from the step, not on top of the clear-cut
line. However, the vortices along the bottom step seem to locate
in both sides of the line. This may suggest that twin boundaries are
induced near the step, which may have weaker superconductivity and
thus construct a strong pinning well. This is similar to the recent
report of the vortex state near the twin boundaries.\cite{Molertwinboundary}

\subsection{Vortex image in UD sample}\label{UD}

As mentioned in Part~\ref{OP}, there are some vortex chains in
optimally doped sample as shown in Fig.~\ref{fig2}, while the
pinning force per unit length is also large on these vortices, as
shown in Figs.~\ref{fig6}(a),(c). There are some reports on the
vortex chain state in the presence of a tilted magnetic field from
the $c$-axis in cuprates\cite{BSCCOchain,YBCOchain}. It was shown
later that these vortex chains are formed by the vortex pancakes
which are dragged by the underneath Josephson vortex. Therefore to
observe the vortex chains in Bi-2212, we need a significant
misalignment, i.e. more than 45$^\circ$, between the direction of
the magnetic field and the $c$-axis of the sample, while in our
experiment the misalignment is smaller than 5$^\circ$.
Additionally, no evidence for either pancake vortices or the
Josephson vortices was observed in the K-doped 122.

\begin{figure}
\includegraphics[width=8cm]{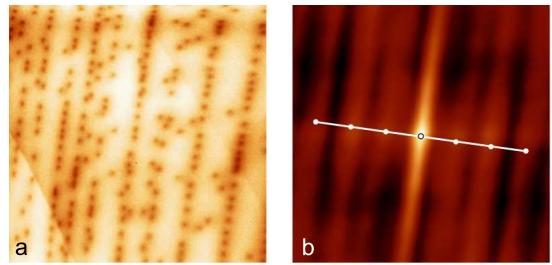}
\caption{(Color online) Vortex image (a) and its self-correlation map (b) of an underdoped Ba$_{0.77}$K$_{0.23}$Fe$_2$As$_2$ sample at 10$\;$Oe and 2$\;$K with a scanning range up to 22$\;\mu$m$\times22\;\mu$m. The white line shows the main distance of the chains, which is about 3.2$\;\mu$m in each small division.} \label{fig8}
\end{figure}

To investigate the vortex chains, we did the further experiment on
the underdoped samples. Fig.~\ref{fig8} shows the vortex image of
the underdoped Ba$_{0.77}$K$_{0.23}$Fe$_2$As$_2$. Clearly, there
are vortex chains along the same direction, and the distances
between the chains are several micrometers. The vortex chains in
underdoped sample are very similar to the ones found in twinned
YBa$_2$Cu$_3$O$_{7-\delta}$\cite{YBCOtwinchain1,YBCOtwinchain2} or
ErNi$_2$B$_2$C\cite{ErNiBC}. The iron pnictides have the
spin-density-wave and the structural transition in underdoped
samples\cite{DaiPCSDW,BaKPhaseD}. This structural transition from
orthorhombic to tetragonal causes the twin boundaries which are
parallel each other and have several micrometers in
distance\cite{domain}. The self-correlation result of these vortex
images shows that the averaged space between the two vortex chains
is about 3.2$\;\mu$m, which is consistent with the space between
the domain walls. No doubt, these twin boundaries enhance the
critical current density as counted from the gradient of the
vortex density.\cite{domainpinning,VortexImage3} If we come back
to the case in optimally doped sample, some strong pinning centers
also show up near the vortex chains, as shown in Fig.~\ref{fig6}.
Therefore, even in the optimally doped sample, we also have the
twin boundaries as the strong pinning centers. It is these strong
and large scale pinning centers that enhance the critical current
density greatly in the weak field region, leading to a sharp
magnetization peak near zero field. When the magnetic field is
increased to a high value, more vortices fill into the area
between the ``network'' of these twin boundaries. Therefore it is
quite natural to observe some complex structure of magnetization
hysteresis loops which exhibit multiple magnetization
peaks.\cite{MultiplePeaks} Our MFM data here give a direct
visualization of this kind of picture for vortex pinning in the
iron pnictide superconductors.

\section{Conclusions}

We present the vortex images on optimally doped Ba$_{0.6}$K$_{0.4}$Fe$_2$As$_2$ and underdoped
Ba$_{0.77}$K$_{0.23}$Fe$_2$As$_2$ single crystals with magnetic fields below
100$\;$Oe. The vortices are very diluted and widespread when the
magnetic field is several oersteds, but they get crowded with a similar
distance at higher magnetic field (higher than 20$\;$Oe). Some vortex
chains are observed together with a roughly random distribution
(with short-range hexagonal order) of vortices between them. The
calculated pinning force per unit length seems much smaller than
that in the Co-doped 122 system at the same field,
indicating that the pinning at Fe sites yields stronger pinning and
vortex disorders. The vortex system becomes more ordered and
favors a six-neighbored structure at higher magnetic field. We
find some surface steps as the pinning centers but they may not act
as the Bean-Livingston mirror pinning. A vortex chain state is also
observed in the underdoped sample and is ascribed to the pinning
by the twin boundaries generated by the structural distortion of
the orthorhombic state. We observed a cooperative pinning induced
by the large-scale twin boundaries and the weak local disorders,
which may be a common picture to describe the vortex dynamics in
iron pnictide superconductors.

\section*{Acknowledgments}
We appreciate useful discussions with Ted Forgan and C. van der
Beek. This work is supported by the NSF of China, the Ministry of
Science and Technology of China (973 Projects: No. 2011CBA00102 and
No. 2010CB923002), PAPD, and the Chinese Academy of Sciences.

$^*$ huanyang@nju.edu.cn

$^\dag$ hhwen@nju.edu.cn

\end{document}